\documentclass[iop]{emulateapj}
\usepackage{color}

\usepackage{epstopdf}
\usepackage{amsmath}
\makeatletter
\newcommand\lsim{\mathrel{\rlap{\lower4pt\hbox{\hskip1pt$\sim$}}
\raise1pt\hbox{$<$}}}
\newcommand\gsim{\mathrel{\rlap{\lower4pt\hbox{\hskip1pt$\sim$}}
\raise1pt\hbox{$>$}}}

\makeatother
\shorttitle{ Dark Matter Torii in BH binaries }
\shortauthors{Naoz \&  Silk}
\makeatother
\begin{document}
\title{ Formation of Dark Matter Torii Around Supermassive Black Holes Via The Eccentric  Kozai-Lidov  Mechanism }

\author{Smadar Naoz\altaffilmark{$\dagger$}} 
\affil{Harvard Smithsonian Center for Astrophysics, Institute for
  Theory and Computation, 60 Garden St., Cambridge, MA 02138\\
  Department of Physics and Astronomy, University of California, Los Angeles, CA 90095, USA}
   \altaffiltext{$\dagger$}{Einstein Fellow}
\email{snaoz@astro.ucla.edu}
\author{Joseph Silk}
\affil{Institut d'Astrophysique de Paris, CNRS, UPMC Univ Paris 06,
UMR7095, 98 bis, boulevard Arago, F-75014, Paris, France \\  The Johns Hopkins University, Department of Physics and Astronomy, Baltimore, Maryland 21218, USA  \\Beecroft Institute of Particle Astrophysics and Cosmology, University of Oxford, Oxford OX1 3RH, UK }

\begin{abstract}
We explore the effects of long term secular perturbations on the distribution of dark matter particles around Supermassive Black Hole (BH) binaries. We show that in the hierarchical (in separation) three-body  problem,  one of the BHs and a dark matter particle form an inner binary. Gravitational perturbations from the BH companion, on a much wider orbit, can cause the dark matter particle to reach extremely high eccentricities and even get accreted onto the BH, by what is known as the Eccentric Kozai-Lidov (EKL) mechanism. We show that this may produce a torus-like configuration for the dark matter distribution around the {\it less} massive member of the BH binary. 
We first consider an Intermediate BH (IMBH) in the vicinity of our Galactic Center, which may be a relic of a past minor merger. We show that if the IMBH is close enough (i.e., near the stellar disk) the EKL mechanism is very efficient in exciting the eccentricity of dark matter particles in near-polar configurations  to extremely high values where they are accreted by the IMBH.  We show that this mechanism is even more effective if the central BH grows in mass, where we have assumed adiabatic growth. Since  near-polar configurations are disrupted, a torus-like shape is formed. We also show that this behavior   is also likely to be relevant for  Supermassive BH binaries.  We suggest that    if the BHs are spinning, the accreted dark matter particles may linger in the ergosphere and thereby may generate self-annihilations and produce  an indirect signature of potential interest.
\end{abstract}

\section{Introduction}\label{intro}

 Observations suggest that   most local galaxies host central supermassive black holes \citep[SMBH, e.g.,][]{Kormendy+95,Richstone+98,Ferrarese+05}.  Thus, within the hierarchical structure formation paradigm, major galaxy mergers may  result in the formation of SMBH binaries \citep[e.g.,][]{DiMatteo+05,Hopkins+06,Robertson+06,Callegari+09}. The evolution of these  binaries is highly dependent on the conditions of the host galaxies.  Numerical experiments for spheroidal gas-poor galaxies suggest that these binaries can reach parsec separation and may stall there \citep[e.g.,][]{Begelman+80,Milosavljevic+01,Yu02}. 

Observations of SMBH binaries are challenging, however  already few systems and several potential candidates have been observed between sub- to few hundreds of parsec separations  \citep[e.g.,][]{Sillanpaa+88,Rodriguez+06,Komossa+08,Bogdanovic+09,Boroson+09,Dotti+09,Batcheldor+10,Deane+14,Liu+14}. Furthermore, several quasar pairs (from which  SMBH pairs are inferred) with kpc-scale  separations have been found in merging galaxies \citep[e.g.,][]{Komossa+03,Bianchi+08,Comerford+09bin,Liu+10,Liu+10kpc,Green+10,Smith+10,Fu+11}. Here we focus on SMBH binaries and explore the effects of their gravitational potential on their Dark Matter (DM) halos.

Another possible configuration that we consider here is a binary consisting of a SMBH and an intermediate mass black hole (IMBH). For this configuration, a low mass  galaxy merging into a more massive galaxy needs to harbor an IMBH and keep its original DM halo. This scenario was considered recently by \citet{RM13} who showed that if seed central Black Holes (BHs)  are common in the first galaxies, then relics of IMBHs should be present in the galactic bulge and halo. They showed that some of these IMBHs may retain  ``their own" DM sub-halos.  In addition,  the possibility of the existence of an IMBH in the central inner parsec of the Milky Way Galaxy has been suggested in the literature due to both observational and theoretical reasons \citep[e.g.,][]{Hansen+03,Maillard+04,GR05,Gualandris+09,Chen+13,RM13}. However, the existence  of such an IMBH is still controversial. For example, \citet{Trippe+08} showed that the dynamics of the old stellar cluster in the Galactic Center is well described by a  relaxed system with no evidence for any large-scale disturbance by an IMBH. In contrast, \citet{Mer+09} showed that the dynamical properties of the S-star cluster are consistent with the nearby existence of an IMBH.   Some constraints on the possible IMBH mass and distance from the galactic center already exists; these exclude  the existence of very massive IMBHs   in a very narrow band of  distances from the central SMBH \citep[e.g.][]{Gualandris+09,Chen+13}.

Here we study the outcome of SMBH gravitational perturbations on the DM particles around a less massive companion (either a SMBH or an IMBH). The SMBH can drive the DM particles to extremely high eccentric orbits until they are plunging into the BH. This may produce a torus-like configuration for the DM particle distribution around the less massive member of the BH binary.

The DM  distribution in the centers of galaxies has  been studied extensively  in the literature. 
Rotation curve analyses of low-surface-brightness spiral galaxies yield low, core-like DM densities in the centers of these galaxies  \citep[e.g.][]{Burkert95,Salucci+00,deBlok+02,deBlok05,Gentile+05}. Although some effects may be missing in these interpretations, such as non-circular motions and gas pressures~\citep[e.g.][]{Simon+05,Spekkens+05,Valenzuela+07} it is likely that such effects  cannot account for significant deviations from core-like density distributions    \citep[e.g.][]{Gentile+05}.  On the other hand, N-body simulations suggest the existence of a density cusp  in the center (e.g., \citet{Dubinski+91}, \citet{Navarro+96,Navarro+97} and see \citet{Kuhlen+12} for a review). These simulations can  determine the halo density profile on scales $0.01 - 1$ of the virial radius of the halo, which is of the order of $\sim 100$~kpc, for a Milky Way-like galaxy. Thus, estimating the density on sub-parsec scales requires extrapolation from these results. If these extrapolations are correct, and because of the   large gravitational potential well in the center, DM particle decays or annihilations may be enhanced. This can lead to a potentially detectable signature despite the uncertain effects of baryonic effects on core scales \citep{Pontzen+12}.

 \citet{GS99} showed that in the presence of an adiabatically growing massive black hole (MBH) at the center of the Galaxy, the density profile can redistribute, forming a spike in the DM density, around the central black hole (MBH). The assumption in this study is that the  MBH accretes mass via spherically-symmetric infall of gas, which would lead to an  increase of the density of the matter around it \citep{Peebles72,Young80}.  This may  also lead to an enhanced rate of DM particle decays or annihilations\footnote{Note that the old stellar density profile would respond similarly to the DM profile.}. It is interesting to note that stellar heating would soften any DM profile at the galactic center, to  an expected $\sim r^{-3/2}$ profile  as inferred for the old stars near the galactic center. However, considering, for example the case of M87, stellar dynamical  heating is ineffective at the lower DM density and higher velocity dispersion in the core, hence the DM spike remains \citep{Vasiliev+08}.

In recent years, the prospect of detecting a signature of DM  has captured the community's interest  \citep[e.g.][]{Bergstrom+98,Ullio+01,Merritt+02,Bertone+05,BM05,Bertone+07,Bertone+09,Hooper+11,Bringmann+12,Weniger12,Buchmuller+12,Daylan+14}.  Specifically, the recent claims for the  $130$~GeV line-like feature \citep{Bringmann+12}  and the excess emission at $1-3$ GeV energies may be interpreted as the signature of DM annihilations in the inner parts of the Galaxy  \citep{Hooper+11,Su+12line,Finkbeiner+13,Daylan+14}.  

 Here we examine the DM density profile in the case of BH binaries. While supermassive black hole (SMBH) binaries are an expected consequence of galaxy mergers, any intermediate black hole (IMBH) in the vicinity of the Galactic Center can also be considered as a binary, and thereby  result in  interesting effects.  We show that secular (i.e., coherent interactions on time-scales long compared to the orbital period) effects play an  important role in shaping the DM density profile.  Specifically these effects may result in a torus-like configuration around the less massive binary member.
 
 We first review the  relevant secular dynamic physics for this configuration \S \ref{sec:EKL}. We then apply this to a BH-IMBH binary in the center of our galaxy \S \ref{sec:IMBH} and then show the effects while considering a growing MBH \S\ref{sec:growMBH}. We then show that long-term evolution also plays an important role for SMBH binaries \S \ref{sec:growMBH}. We discuss some of the implications  of our results in \ref{sec:Dis}.

\section{Long-Term Dynamical Evolution}\label{sec:EKL}

\begin{figure}[!t]
\includegraphics[width=\linewidth]{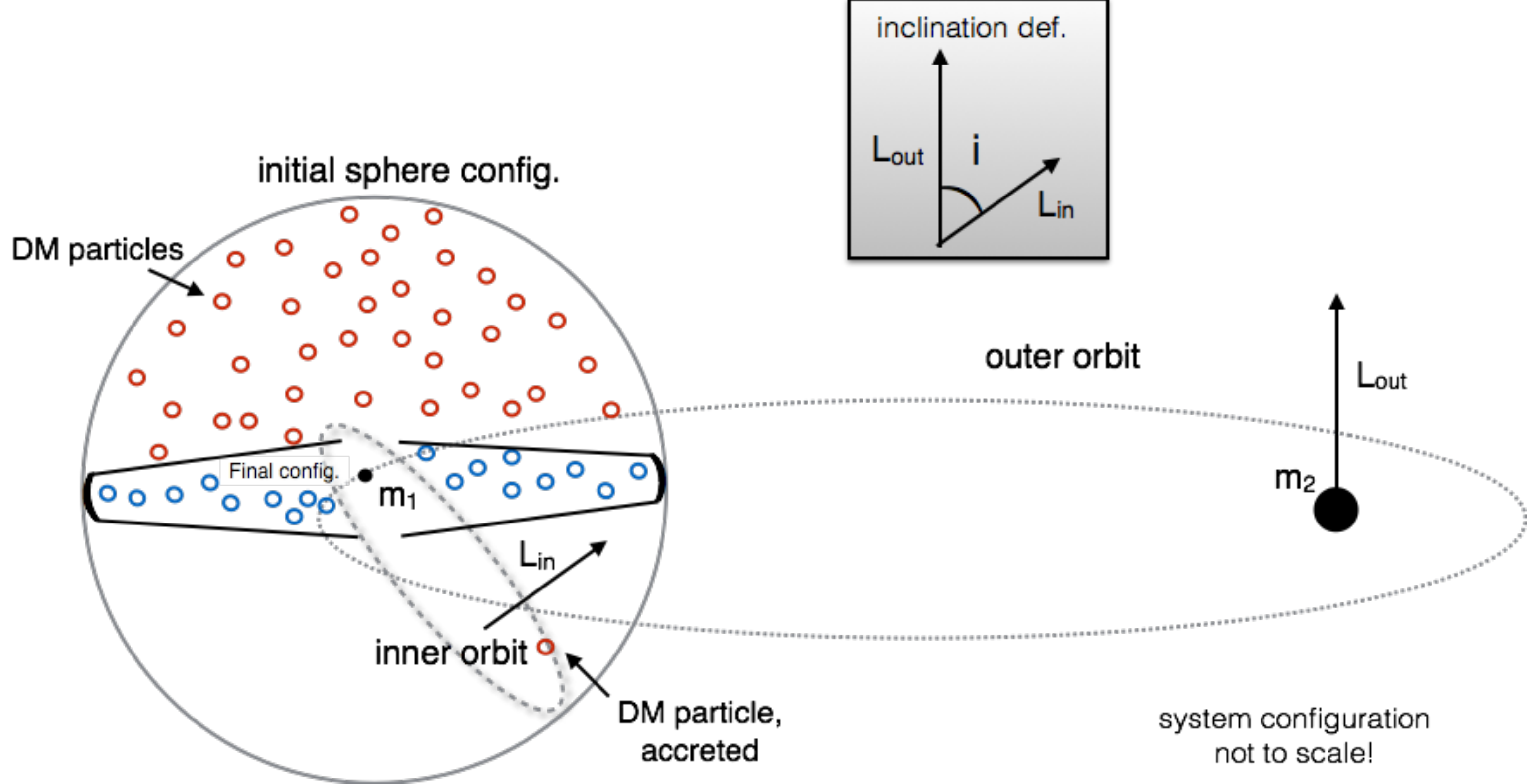}
\caption{Cartoon description of the configuration considered here. Each black hole starts with a sphere of DM particles (the sphere around $m_2$ not shown in the figure). The particles in a near-polar orbit with respect to the BH binary orbit, will undergo large eccentricity and inclination oscillations.  The grey box shows the definition of the inclination $i$, where $\cos i={\bf L}_{tot} \cdot {\bf L}_{in}$}
   \label{fig:config} 
\end{figure}

We consider the long-term gravitational interactions of a binary BH on the surrounding DM particles (see Figure \ref{fig:config} for a cartoon configuration). We study this configuration in the framework of a hierarchical (in separation) three-body system.  Here, dynamical stability requires  a hierarchical configuration, in which the {\it inner binary} - in our case the one consisting of the SMBH and the DM particle (with masses $m_1$ and  $m_{DM}\to 0$, respectively and separation of $a_{in}$), see Figure \ref{fig:config} -  is orbited by a third body, the second SMBH with mass $m_2$, on a much wider orbit,  i.e.,  the {\it outer binary}. The outer orbit  is at a separation  $a_{out}\gg a_{in}$. The two orbits can be eccentric ($e_{in}$ and $e_{out}$, for the inner and outer orbits, respectively), and inclined with respect to one another, defined as the angle between the two  angular momenta (see grey box in Figure  \ref{fig:config}). 
In this case the secular approximation (i.e., phase averaged, long term evolution) can be applied, where the interactions between the two  non-resonant orbits is equivalent to treating the two orbits as massive wires. Here the line-density is inversely proportional to orbital velocity and the potential is being expanded in semi-major axis ratio (which in this approximation, remain  constant), $a_{in}/a_{out}$, \citep{Kozai, Lidov}.  This ratio  is a small parameter due to the hierarchical configuration. The  lowest order of approximation, which is proportional to $(a_1/a_2)^2$  is called quadrupole level.

In early studies of hierarchical secular three-body systems  
\citep{Kozai,Lidov}, the outer orbit  was set to be circular and   one of the inner binary members  is assumed to be a test (massless) particle.  In this situation, the
component of the inner orbit's angular momentum along  the total angular momentum  is
conserved, and the lowest order of the approximation, (i.e., the quadrupole approximation), is valid.  \citet{Kozai} and \citet{Lidov}  found that  for initial large mutual inclinations  between the inner and outer orbit ($40^\circ \lsim i \lsim 140^\circ$, see grey box in Figure \ref{fig:config}) and initial small inner orbit eccentricity, 
the inner orbit's eccentricity and inclination can have large amplitude oscillations.

 Recently, \cite{Naoz11,Naoz+11sec} showed that considering systems beyond the test particle approximation, or  circular orbits, requires the next level of approximation for a correct representation of the physics, called the octupole--level, which is proportional to $(a_1/a_2)^3$. In this case,  the angular momenta component of the inner and outer orbits  along  the total angular momentum is not conserved. 
In the octupole level of approximation, the inner orbit eccentricity can reach very high values \citep{Ford00,Naoz+11sec,Li+13,Li+14,Tey+13}.  Furthermore, the inner orbit inclination can flip its orientation from prograde, with respect to the total angular momentum, to retrograde \citep{Naoz11,Naoz+11sec}.  We refer to this process as the {\it eccentric Kozai--Lidov} (EKL) mechanism.

It has been shown that the Kozai-Lidov mechanism (and specifically the EKL) has important implications in different astrophysical settings, such as triple stars \citep[e.g.,][]{Har69,Mazeh+79,1998KEM,Dan,PF09,Tho10,Naoz+11sec,Prodan+12,Sharpee+12,Prodan+13,Naoz+14} and extrasolar planetary systems with an additional distant 
companion \citep[e.g.,][]{Hol+97,Dan,Wu+07,Takeda,Naoz11,Naoz+12bin,Li+13,Tey+13,Petrovich+14}. In addition, the Kozai-Lidov mechanism has been suggested as  playing an important role in both the growth of black holes (BHs) at the centers of dense stellar clusters and the formation of short-period binary BHs \citep{Bla+02,MH02,Wen}.  Furthermore, \citet{Iva+10} suggested that the most important formation channels  for BH X-ray binaries in globular clusters may be triple-induced mass transfer in a BH-white dwarf binary.

We solve the octupole--level approximation of the hierarchical three-body problem \citep[see][]{Ford00,Naoz+11sec}, including the $1st$ post-Newtonian general relativity effects \citep[see][]{Naoz+12GR}. We restrict ourselves to the test particle approximation since the DM particle mass is negligible. The relevant octupole level of approximation equations and further discussion of this approximation can be  found in \citet{LN11}, and \citep[see also][]{Boaz2,Li+14}.

The contribution of the different physical processes that  affect the evolution of the system can be estimated by considering the time-scales of the precessions  by the different mechanisms.
The time-scale associated with the precession of the inner orbit due to the quadrupole term can be estimated from the canonical equations of motion \citep[e.g.][]{Naoz+12GR}, for a test particle 
\begin{equation}\label{eq:tquad}
t_{\rm quad}\sim \frac{2\pi a_{out}^3 (1-e_{out}^2)^{3/2}\sqrt{m_1}}{ a_{in}^{3/2} m_2 \sqrt{G}}  \ ,
\end{equation}
where $G$ is the gravitational constant. 
The time-scale associated with the  octupole term is more difficult to estimate due to its chaotic behavior, however, following  \citet{Naoz+12GR} we give a rough estimate, for the regular part of the evolution  in the form of  \citep[e.g.][]{LN11}
\begin{equation}\label{eq:toct}
t_{\rm oct}\sim \frac{4}{15} \frac{1}{\epsilon} t_{\rm quad} \ ,
\end{equation}
  for a given inner and outer eccentricity, 
  \begin{equation}\label{eq:epsilon}
  \epsilon=\frac{a_{in}}{a_{out} } \frac{e_{out}}{1-e_{out}^2} \ .
  \end{equation}
  The $\epsilon$ parameter in Equation (\ref{eq:epsilon}) is the pre-factor before the octupole level of approximation of the Hamiltonian \citep[e.g.,][]{Naoz+11sec,LN11} furthermore it also can be used as a stability condition (we discuses the stability of the system more below). 
  Note that the more general octupole time-scale depends not only on $e_{in}$ but also on the mutual inclination \citep[e.g.][]{Li+13,Tey+13}. However, the octupole time-scale in equation (\ref{eq:toct}) provides a sufficient limiting criterion for the regular high inclination behavior \citep{Naoz+12GR}.  The octupole level time-scale gives the relevant time-scale to reach extremely large eccentricities. To accrete a DM particle onto a BH, the DM orbit needs to reach an eccentricity of almost unity, and thus the octupole level of approximation plays an important role.   
 
\begin{figure}[!t]
\includegraphics[width=\linewidth]{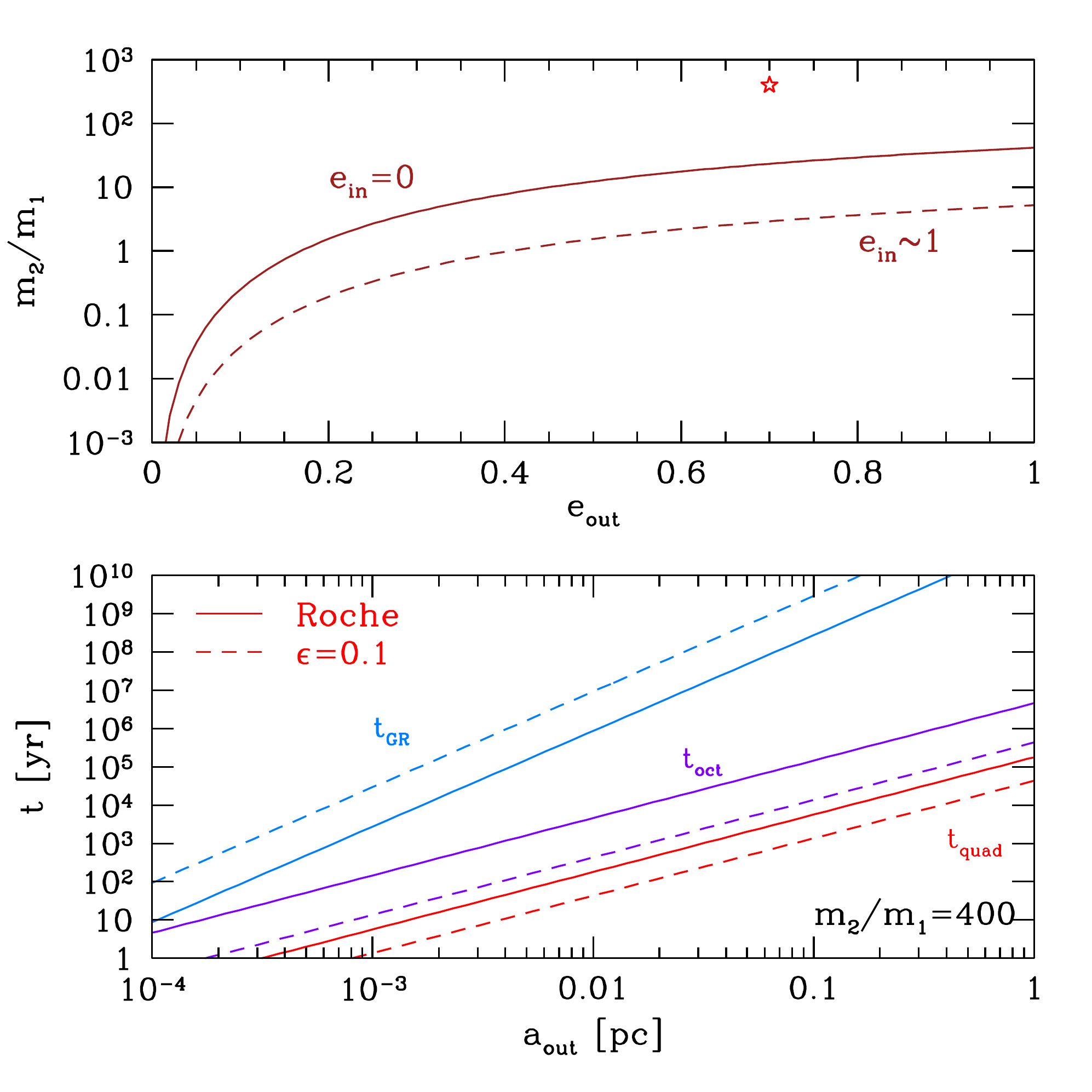}
\caption{Stability requirements for a hierarchal system and the dynamical time-scales that affect DM particles. {\bf Top panel:} The stability requirements derived from equation (\ref{eq:Stab}) as a function of the binary BH eccentricity ($e_{out}$), for zero inner orbit eccentricity (solid line) and large eccentricity $e_{in} \to 1$ (dashed line). During the evolution, the inner eccentricity oscillates and can reach extremely large values. Thus the dashed line gives the lower stability limit. The red star marks the location of $m_2/m_1=400$ and $e_{out}=0.7$, which are the parameters assumed for the bottom panel.  {\bf Bottom panel,} shows the relevant time-scales of the evolution as a function of the outer orbit separation $a_{out}$, for a system with $m_2/m_1=400$ and $e_{out}=0.7$. We consider the quadrupole time-scale, Eq.~(\ref{eq:tquad}), the octuple time-scale, Eq.~(\ref{eq:toct}) and the inner orbit GR precession time scale Eq.~(\ref{eq:tGRa1}), red, purple and blue lines, respectively. For each of these time-scales, we choose two possible inner orbit separations. The first (dashed lines) satisfy the requirement of $\epsilon=0.1$, while the second (solid line) is the Roche limit [Eq.~(\ref{eq:Roche})], for zero inner orbit eccentricity.
}
   \label{fig:timescales} 
\end{figure}

  The DM particle orbit around a BH will  also precess due to  general relativity effects. The time-scale of this effect can be estimated as  \citep[e.g.,][]{Naoz+12GR} :
\begin{equation}
\label{eq:tGRa1}
t_{GR}\sim 2\pi \frac{a_{in}^{5/2} c^2 (1-e_{in}^2) }{ G^{3/2} (m_1)^{3/2}} \ ,
\end{equation}
where $c$ is the speed of light.
This precession is taking place in the opposite direction from that which is dictated by the EKL mechanism. Therefore, if the GR precession  time-scale  is shorter than $t_{quad}$, eccentricity excitations may be suppressed. However when the two time-scales are similar, the eccentricity may be excited to large values in a resonance-like behavior  \citep{Naoz+12GR}. Furthermore, as pointed out by \citet{Antognini+13} these systems and packed systems may result in even larger eccentricity excitations due to GR effects \citep[see also][]{Will13}. 
We show these time-scales (Equations (\ref{eq:tquad}), (\ref{eq:toct}) and (\ref{eq:tGRa1})) in the bottom panel of Figure \ref{fig:timescales} as a function of the outer orbit separation $a_{out}$, for $m_2/m_1=400$, $e_2=0.7$.

The inner orbit will also precess  due to the enclosed  mass inside the DM particle orbit, the larger the mass the faster  the precession \citep[e.g.][]{Tremaine05}. Similarly to the GR precession, this precession takes place in the opposite direction from the EKL and thus may suppress eccentricity excitations.  
It is hard to estimate the mass inside the sphere of influence of a black hole. Estimates from simulations are difficult due to the different dynamical scales \citep[e.g.,][]{RM13}. Therefore, we extrapolate a core density (i.e., constant density) to the innermost radii from the upper limits mentioned in  \citet{Merritt10}. The  choice of a core (rather than cusp) distribution is supported by  the \citet{Quinlan+97} study that showed that BH binary systems may result in a core distribution of both the DM and stars. 
 Below, we consider a configuration of DM particles around $10^4$~M$_\odot$ BH (see Figures \ref{fig:IMBH} - \ref{fig:IMBHgrow}). For this configuration and relevant scales (see below), we adopt a density of  $1$~M$_\odot$~pc$^{-3}$ from   \citet{Merritt10}, as an upper limit and  we find that the Newtonian time-scales are irrelevant for these configurations (larger than Hubble time). We therefore  omit this time-scale from Figure 
\ref{fig:timescales}. We discuss this physical effect below for the different configurations.

With the time-scales specified above, we now analyze the parameter space. This will yield the configurations that are most likely to be affected by the EKL mechanism. 
The first condition is hierarchy in separations, where, as mentioned, 
the $\epsilon$ parameter in Equation (\ref{eq:epsilon})  can be used to determine hierarchy. 
A system is considered hierarchical as long as $\epsilon <0.1$, where systems above this value are more likely to become unstable \citep[e.g.,][]{LN11}. For a given mass ratio and $e_{out}$ we find the maximal separation ratio $a_{in}/a_{out}$ that a stable system has. In   Figure \ref{fig:timescales} bottom panel, we show the relevant time scales of this maximal separation (dashed curves).

\begin{figure*}[!t]
%\hspace{-0.7cm}
\begin{center}
\includegraphics[width=0.9\linewidth]{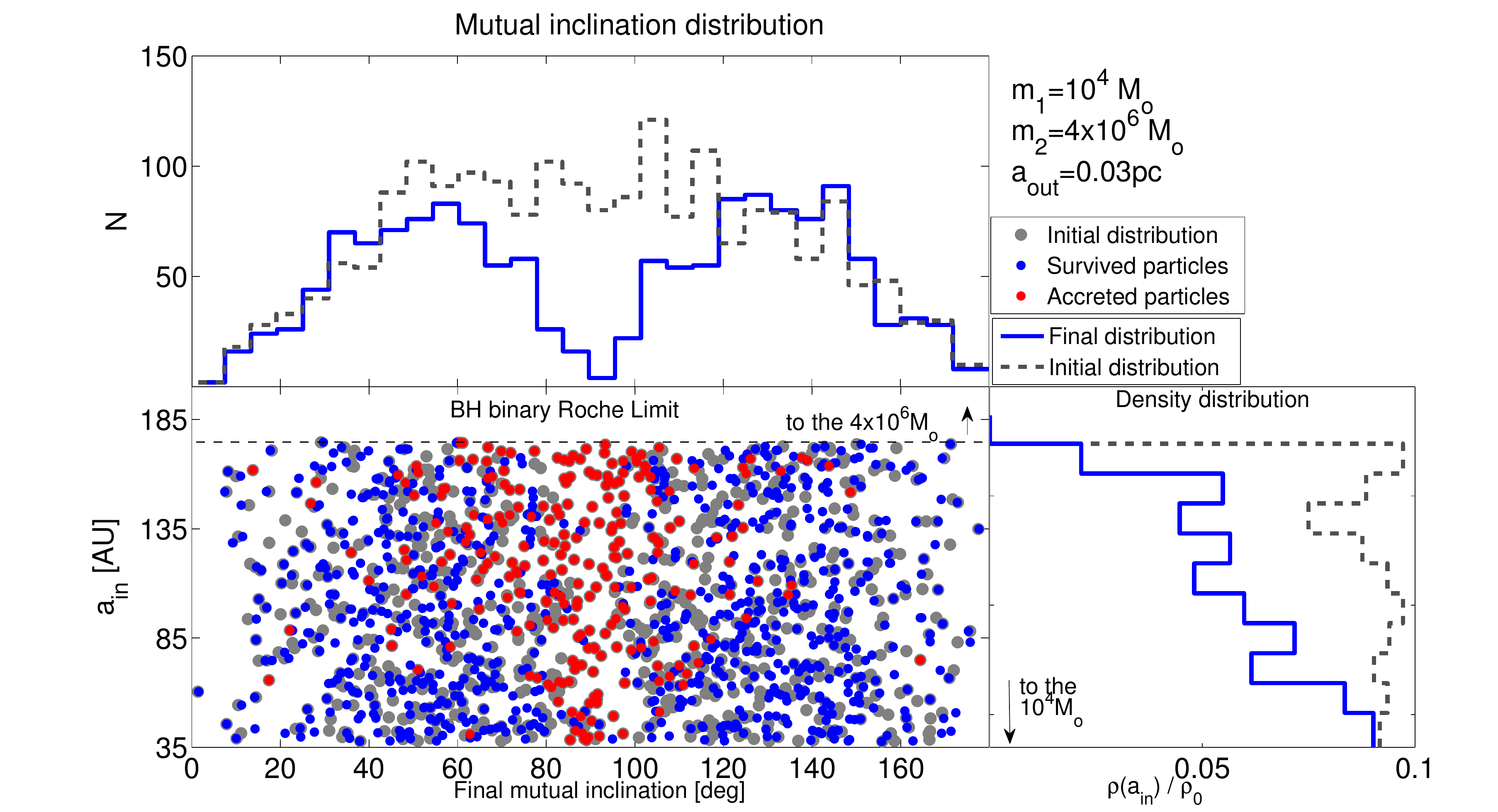}
\caption{{\bf The final DM particles distribution around the IMBH after $10^8$~yr of evolution.} We consider a  IMBH with $10^4$~M$_\odot$ at $0.03$~pc from the central $4\times10^6$~M$_\odot$ MBH. {\it The top panel} shows the initial isotropic distribution (grey dashed line) and the final inclination distribution (blue solid line).  {\it The bottom left panel}  shows the DM particle separations ($a_{in}$, in AU since the particles are very close to the IMBH) vs. the mutual inclination. We show the initial distribution of the particles (light large grey dots), on top of them we mark in red  those configurations that resulted in pericenters smaller than $r_{c}$. We also overplot here the final mutual inclination of the particles that survived (blue dots). The dashed line shows the Roche-limit separation for circular orbits [Eq.~\ref{eq:Roche}] . {\it The bottom right panel} shows the DM particles distribution normalized to have an integral of unity. We consider the initial distribution (grey dashed line), and the final DM distribution of the particles that survived (blue solid line).
 The plot shows the result of $1974$ runs.
}
   \label{fig:IMBH} 
   \end{center}
\end{figure*}

During the BH binary orbit, DM particles around one of the binary members may be tidally attracted to the other member. The radius at which this happens, also known as the Roche limit, is estimated as:
\begin{equation}\label{eq:Roche}
a_{in}(1+e_{in}) \sim a_{out} (1-e_{out}) \left(\frac{m_1}{m_2}\right)^{1/3} \ .
\end{equation}
DM particles around $m_1$ with larger separations will feel a larger gravitational force from $m_2$. From this equation and equation (\ref{eq:epsilon}), we can find the mass ratio that will result in a stable configuration as a function of the BH binary mass ratio, i.e., 
\begin{equation}\label{eq:Stab}
\frac{m_2}{m_1}=\frac{1}{3}\left(\frac{e_{out}}{\epsilon (1+e_{in})(1-e_{out})}\right)^3 \ .
\end{equation}
Setting the stability requirements of $\epsilon=0.1$, gives a lower limit for the mass relation. We show this relation in the top panel of figure \ref{fig:timescales}. This means that the EKL mechanism may have a significant effect for eccentric BH binaries where the perturber has the larger mass.  Therefore, in the configurations considered below, we choose $m_2>m_1$ and we focus on the DM particle distribution around $m_1$.  For example,  in the cartoon in Figure \ref{fig:config} $m_2>m_1$.
Note that since the octupole level of approximation cancels out for $e_{out}\to 0$, \citep{Naoz+11sec},  we limit ourselves to eccentric binaries.

The bottom panel of Figure \ref{fig:timescales} shows the relevant time-scales as a function of the outer orbit semi-major axis ($a_{out}$) for a specific mass relation, i.e., $m_2/m_1=400$, and $e_{out}=0.7$.  There are two extreme cases of the inner orbit separation  ($a_{in}$)  depicted in this figure. The first is the one that obeys $\epsilon=0.1$ and the second is the one for which equation  (\ref{eq:Roche}) is satisfy. As can be seen in the bottom panel of Figure  \ref{fig:timescales}, BH binaries at separations $\lsim1$~pc and $\gsim 10^{-4}$ may result in excitation of the inner binary eccentricity  to unity  during the lifetime of the system. Below we explore several examples of the relevant configurations.

\section{DM particles around IMBH}\label{sec:IMBH}

We assume an IMBH of  $10^4$~M$_\odot$, around the central MBH ($4\times 10^6$~M$_\odot$) with an eccentricity of $e_{out}=0.7$. The arguments of pericenter and longitude of ascending nodes were chosen  from a uniform distribution. We choose an isotropic distribution for the DM particle inclinations and a uniform distribution for their eccentricity.  We explore two examples; in the first one the IMBH is located in the stellar disk at the center of the Galaxy, i.e., $a_{in}=0.03$~pc (see Figure \ref{fig:IMBH} ), and in the second example, the outer binary separation is $1$~pc   (see Figure \ref{fig:IMBH1pc}). We chose these values as two representative examples, they are independent of any evidence that IMBH may or may not exists at these distances.  {We speculate that there is a core density distribution of DM particles around the IMBH. This speculation  is supported by the  \citet{Quinlan+97} study of BH binary systems, that suggested a core distribution of both the DM and stars may be a result of dynamical interactions of these binary systems. The upper limit inner orbit separation was chosen to be inside the Roche limit [Eq.~(\ref{eq:Roche})],  where, as implied from Figure \ref{fig:timescales}, the orbit is always stable  (i.e., $\epsilon<0.1$). The two cases yield a similar maximal $\epsilon\sim0.038$. For practical reasons, we also choose a lower limit for $a_{in}$ to be the separation at which the GR time-scale is equal to the quadrupole time scale (i.e., $t_{quad}\sim t_{GR}$), since below this limit, eccentricity excitations are suppressed \citep[e.g.,][]{Dan,Naoz+12GR}.  Guided by the bottom panel of Figure \ref{fig:timescales}, we integrate the systems for $10^8$~yrs for the $a_{out}=0.03$~pc and for $10^9$~yrs for the $a_{out}=1$~pc.   We stop the integration when a particle reaches a pericenter of $r_c=4m_1 G/c^2$. This value is somewhat arbitrary and was chosen to be consistent  with \citet{GS99}. For $10^4$~M$_\odot$ IMBH this value is $3.95\times 10^{-4}$~AU ($=1.92\times 10^{-9}$~pc). 

Stars in the bulge are expected to scatter the DM particles  \citep[e.g.,][]{Vasiliev+08}, as well as produce  an additional source for the Newtonian precession. However,  if the IMBH is the relic of the first galaxies and halos in the early Universe,  their baryon fraction and thus the number of stars may be low 
\citep[e.g.][]{Naoz+11cosmo,Peirani+12,Naoz+13cosmo}. Thus, scattering, or heating of DM particles due to stars is negligible in this case. We note that for the configurations considered here, the Newtonian precession becomes comparable to the EKL timescales  only if the IMBH retains a very large DM density  of about $10^9$~M$_\odot$~pc$^{-3}$ close to the IMBH ($\lsim 200$~AU). As these densities are unlikely, we can safely ignore the contribution of the Newtonian precession in our calculations below.

In Figure \ref{fig:IMBH}  we consider the $a_{out}=0.03$~pc case after $10^8$~yr of evolution. As depicted in this figure,  the DM particles around the IMBH  are redistributed due to the EKL mechanism with a conical void almost perpendicular to the binary MBH-IMBH orbit, and peaks around $\sim 40^\circ$ and $\sim 140^\circ$ (known as the Kozai angles). These angles are associated with the quadrupole level separatrix \citep[e.g.][]{Morbidelli+91,LN11,Li+14}. 
 In the top panel of this figure,  we  show the initial inclination distribution of all DM particles (set to be isotropic, i.e,. uniform in $\cos i$), and the final distribution of the particles that survived, where the two peak distribution yields a torus-like configuration. The particles in the near polar configuration have crossed the $r_c$ limit and we assumed they were accreted onto the IMBH. About $21\%$ of  all runs in that configuration had a pericenter smaller  than the  $r_c$ limit .

In the bottom left panel of  figure  \ref{fig:IMBH}, we show the initial separations of the DM particles ($a_{in}$, which remain constant in the secular  approximation) and the inclinations of DM particles (grey dots). We mark in red those configurations that crossed the  $r_c$ limit (those particles, assumably accreted by the IMBH, and thus they do not reach a final stable configuration). We also over-plot the final configuration of the particles that  survived   (blue dots). The resulted distribution as a function of the DM separation from the IMBH is shown in the right bottom panel. We consider the initial distribution (chosen to be uniform), and the final distribution of the survived particles. 

Closer to the IMBH, particles need to be excited to somewhat less eccentric orbits ($1-e_{in}\sim 10^{-5}$ at our lower limit, compared to the $2\times 10^{-6}$ from the edges of the Roche sphere). However, most of the particles that were accreted by the IMBH arrived from larger distances (as depicted in Figure \ref{fig:IMBH}  bottom panels). This is because these configurations have larger $\epsilon$, where the EKL mechanism is more effective. \citet{Li+14,Li+13}  showed that this eccentricity range is easy to achieve in the test particle approximation if the EKL mechanism is efficient. Most of the particles that were accreted, reached their high  eccentricities  in less than $\sim 3\times 10^5$~yrs of evolution\footnote{Note that this time-scale does not directly relates to the flip time-scale (roughly estimated from the octupole 
time-scale) since we require extremely large eccentricities values}, as depicted in Figure \ref{fig:IMBHmaxe}. Overall after $10^8$~yrs, this mechanism  yields a torus-like configuration with a diluted DM density further away from the IMBH.

\begin{figure}[!t]
\hspace{-0.7cm}
\includegraphics[width=10cm,clip=true]{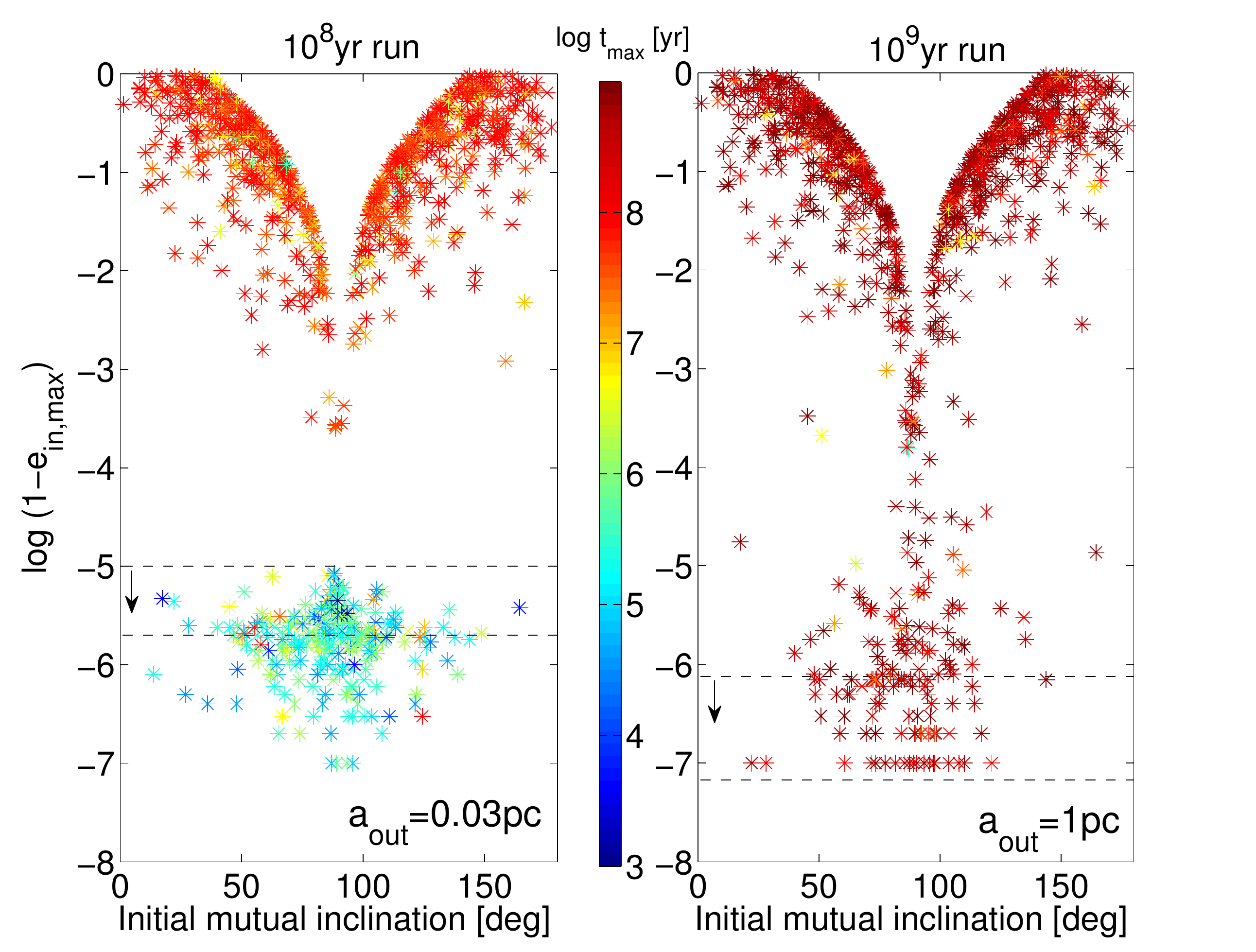}
\caption{{\bf The maximum eccentricity (presented as $1-e_{in}$) reached during the evolution as a function of the initial inclination}.  We consider a  IMBH with $10^4$~M$_\odot$ at $0.03$~pc {\it left panel} and $1$~pc {\it right panel} from the central $4\times10^6$~M$_\odot$ MBH. The different colors indicate the time in years that the maximal value was recorded. Note that about $1\%$ of the runs (in both cases) reached $e_{max}=1$ and are not seen in this Figure. These systems represent about $80\%$ from all the systems that crossed the $r_c$ limit in the $a_{out}=1$~pc case. The horizontal lines are the maximum eccentricities (minimum $1-e_{in}$) the DM particles need to achieve to cross the $r_c$, for the upper and lower limit of the DM initial sphere. For the $a_{out}=0.03$~pc case  (left panel) these are  $2\times 10^{-6}$ and $10^{-5}$, respectively. While for the  $a_{out}=1$~pc case (right panel) these are   $6.8\times 10^{-8}$ and $7.6\times 10^{-7}$. Integration was stopped when $1-e_{in}$ reached the $r_c$ value. See Figures  \ref{fig:IMBH} and \ref{fig:IMBH1pc} for the corresponding separation and final mutual inclination distributions.
}\vspace{0.7cm}
   \label{fig:IMBHmaxe} 
\end{figure}

\begin{figure}[!t]
\hspace{-0.7cm}
\includegraphics[width=10cm,clip=true]{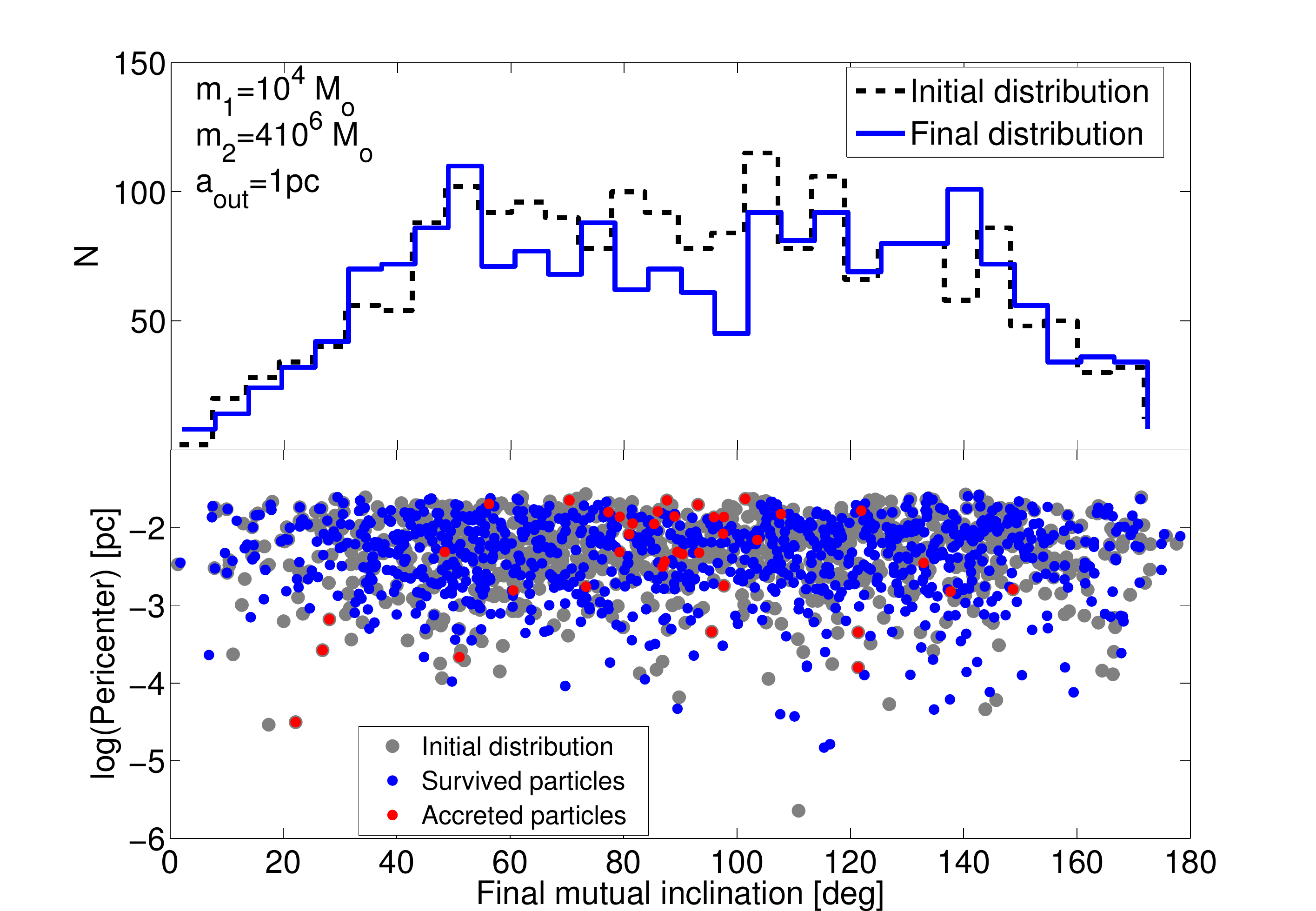}
\caption{{\bf The final DM particles distribution around  the IMBH after $10^9$~yr of evolution.} We consider a  IMBH with $10^4$~M$_\odot$ at $1$~pc from the central $4\times10^6$~M$_\odot$ MBH. {\it The top panel} shows the initial isotropic distribution (grey dashed  line) and the final inclination distribution (blue solid line).  {\it Bottom panel} shows the final pericenter (in pc) as a function of the final inclination (blue dots); we also show the initial pericenter distribution as a function of the {\it initial} inclination in blue,  and we marked in red those configuration that resulted in pericenters smaller than $r_c$. The plot shows the result of $1975$ runs.
}\vspace{0.7cm}
   \label{fig:IMBH1pc} 
\end{figure}

In Figure \ref{fig:IMBH1pc}, we consider the case of an IMBH-MBH binary separated at $1$~pc. We integrate  this configuration  up to 1~Gyr, (see Figure \ref{fig:timescales}, for the reason for the different integration time). Here (Figure \ref{fig:IMBH1pc}, bottom panel) we show the pericenter value (which will better clarify the underlying behavior for this example than the just the separation) as a function of the mutual inclination. Note that in this case parsec  units are more suitable than AU. In this configuration, the inner DM orbits that would be effected from the EKL mechanism (i.e., those for which $t_{quad}>t_{GR}$) are more than an  order of magnitude further away from the $a_{out}=0.03$~pc case ($\sim 520$~AU, compared to the $\sim 37$~AU, respectively). Furthermore, the  Roche limit  is also further away from the IMBH ($\sim 5820$~AU, compared to the $\sim 177$~AU in the $a_{out}=0.03$~pc case). Therefore, the eccentricity that DM particles need to achieve in order to cross the $r_c$ limit is much higher, and ranges between $1-e_{in}\sim7.6\times 10^{-7} - 6.8\times 10^{-8}$.  

Closer to the Roche limit, where $\epsilon$ is closer to its maximal value, and the EKL mechanism is most efficient, the eccentricity that the needs to be reached for a particle to be accreted is extremely large and thus overall this mechanism is less effective in accreting particles onto IMBH, with only $\sim4\%$ of the particles being accreted. Increasing the integration time by a factor of $\sim 10$ will allow the system to explore larger parts of the parameter space and potentially slightly increase the maximal eccentricity (and thus the  fraction of accreted particles will slightly increase). Note that the test particle EKL is well characterized by two parameters; $\epsilon$ and the system energy,   \citep[e.g.][]{LN11}, and the energy of the system is determined by the initial values of $e_{in},e_{out},i$ and the argument of periapsis, which were chosen from the same distribution for the two cases.  Therefore, the $a_{out}=0.03$~pc represents a rescaled version of the  $a_{out}=1$~pc case. Thus, since the former did not produce large numbers of systems with eccentricity exceeding $1-e_{in}<10^{-7}$, (see Figure \ref{fig:IMBHmaxe}, where only about $1\%$ of all  runs have reached $e_{in,max}=1$ and are not depicted in this figure), we do not expect that increasing the integration time for the $a_{out}=1$~pc, will significantly produce more accreted particles.

The inclination distribution in this case, shows less deviation from an isotropic distribution, simply because the near-polar configuration is not significantly  depleted. These configurations are a natural consequence of the EKL mechanism, as shown for example in \citet{Tey+13}, Figure 15 (top left panel ) for $e_{out}=0.7$. The two-peak inclination distribution between the Kozai angles in the $a_{out}=0.03$~pc is a result of the depletion of the near-polar configurations.

The extremely large eccentricities of the DM particles  imply that in some cases the double average approximation used here breaks down, where the angular momentum of the DM particle changes on timescale similar to its orbit timescale \citep{Katz+12,Antognini+13,Antonini+14,Bode+14}.  This  does not change our result of a torus-like final configuration, it simply means that  those systems are poorly described by our approximation and their eccentricity may become even larger during the orbital evolution \citep[e.g.,][]{Antognini+13,Antonini+14}.

  Note that a possible caveat to this calculation is the assumption that the IMBH eccentricity remains constant throughout the evolution.  However, as shown in \citet{Madigan+12} the evolution of an IMBH's eccentricity in the Galactic Center  can be secular in nature. They showed that     interactions with the stellar environment can cause either circularization or even excite   the IMBH's eccentricity.  This  highly depends on the number of stars inside the IMBH orbit and the initial orientation of the IMBH orbit compared to the other stars.
 If the IMBH orbit will be  circularized, $\epsilon$ will be reduced and thus the above effect  will be suppressed.

\begin{figure}[!t]
\includegraphics[width=\linewidth]{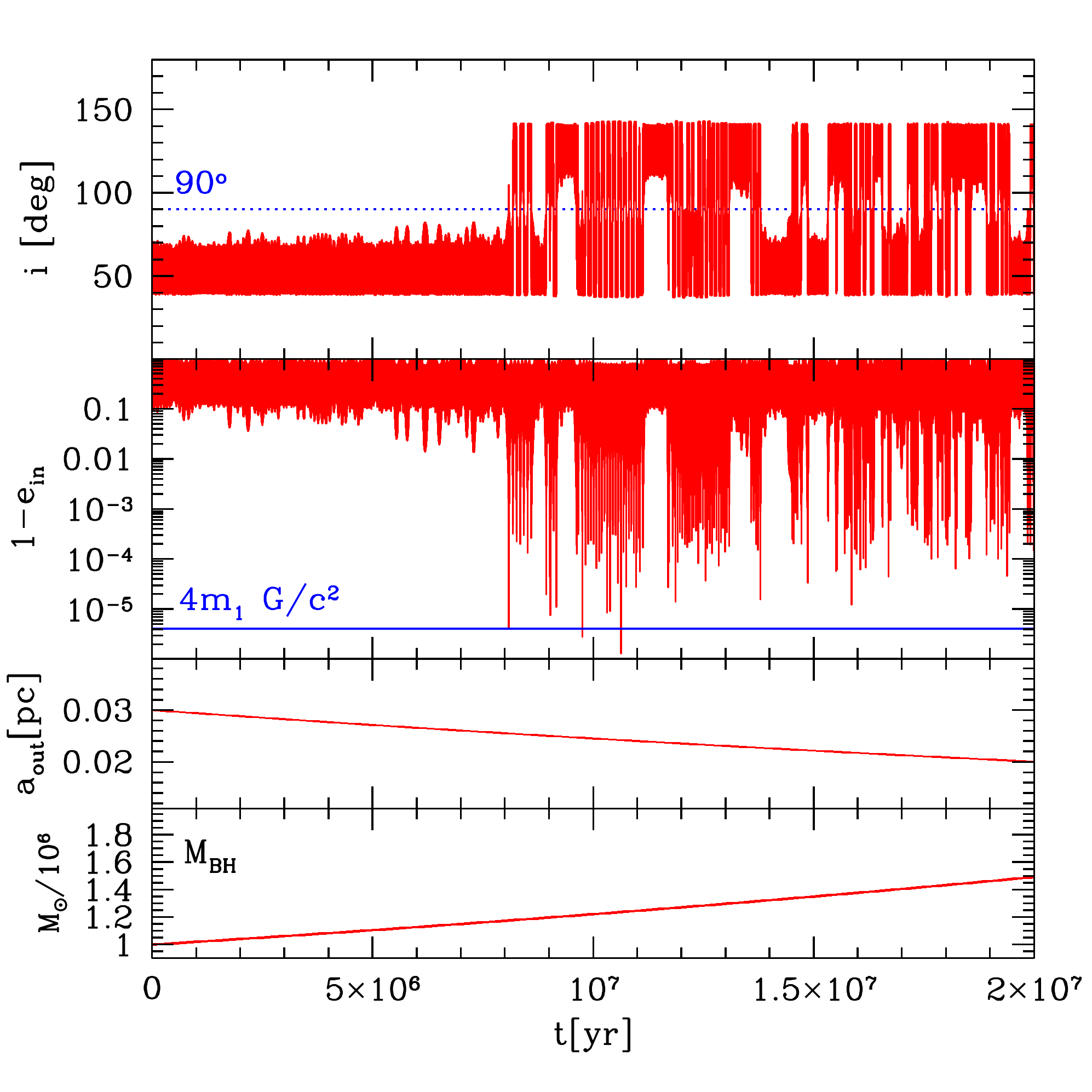}
\caption{An example of the  effect of adiabatic growth of the MBH on the dynamics of the DM particles as a function of time, taken from one of the Monte-Carlo runs shown in Figure \ref{fig:IMBHgrow}.  We consider from top to bottom the orbit mutual inclination, the inner orbit eccentricity (showed as $1-e$) the outer orbit separation, $a_{out}$ and the perturbed mass. The system initial conditions are a $10^4$~M$_\odot$ IMBH with a DM particle at $97.9$~AU. The IMBH is located $0.03$~pc from a MBH set initially on $10^6$~M$_\odot$. The mass of the MBH grows according to Equation (\ref{Eq:BHgrow}).   The other parameters considered for the system are: $e_{out}=0.7$, $e_{in}=0.005$, $i=66.8^\circ$ and the argument of periapsis of the inner and outer orbits were initially set to be $90.13^\circ$ and $155.09^\circ$ respectively. For the given inner orbit separation, we also show the corresponding eccentricity to the $r_c$ limit. For the purposes of this figure, we have continued the evolution beyond this limit.
}
   \label{fig:ExMBHG} 
\end{figure}

 %
%%% up: 4.508359e+02 AU low 1.190256e+03 AU
%%up: 2.185714e-03 AU low 5.770526e-03 pc
%%4M=1.581527e-01 AU and 7.667460e-07 pc
%%1pc : low = 523.2936 AU  up=5.8171e+03 AU
%% 0.03pn low~37AU high ~174AU

\subsection{Adiabatic growth of MBH}\label{sec:growMBH}

\citet{GS99} considered the effect of a growing MBH on the DM distribution around the MBH. They showed that this growth will result in an enhancement of the DM density close to the MBH.  Here we also consider the effects of the growing {\it massive} black hole 
on the DM distribution around the IMBH. We consider a Salpeter MBH growth with the following relation:
\begin{equation}\label{Eq:BHgrow}
M_{BH}(t)=M_{MB,0}(e^{t/t_{ST}} )\ ,
\end{equation}
where $M_{MB,0}$ is the initial black hole mass, and $t_{ST}=5\times 10^7$~yr is the typical growth time. Note that we do not expect the IMBH to grow significantly in mass  since its surrounding is most likely deprived of gas.

\begin{figure}[!t]
\hspace{-0.7cm}
\includegraphics[width=10cm,clip=true]{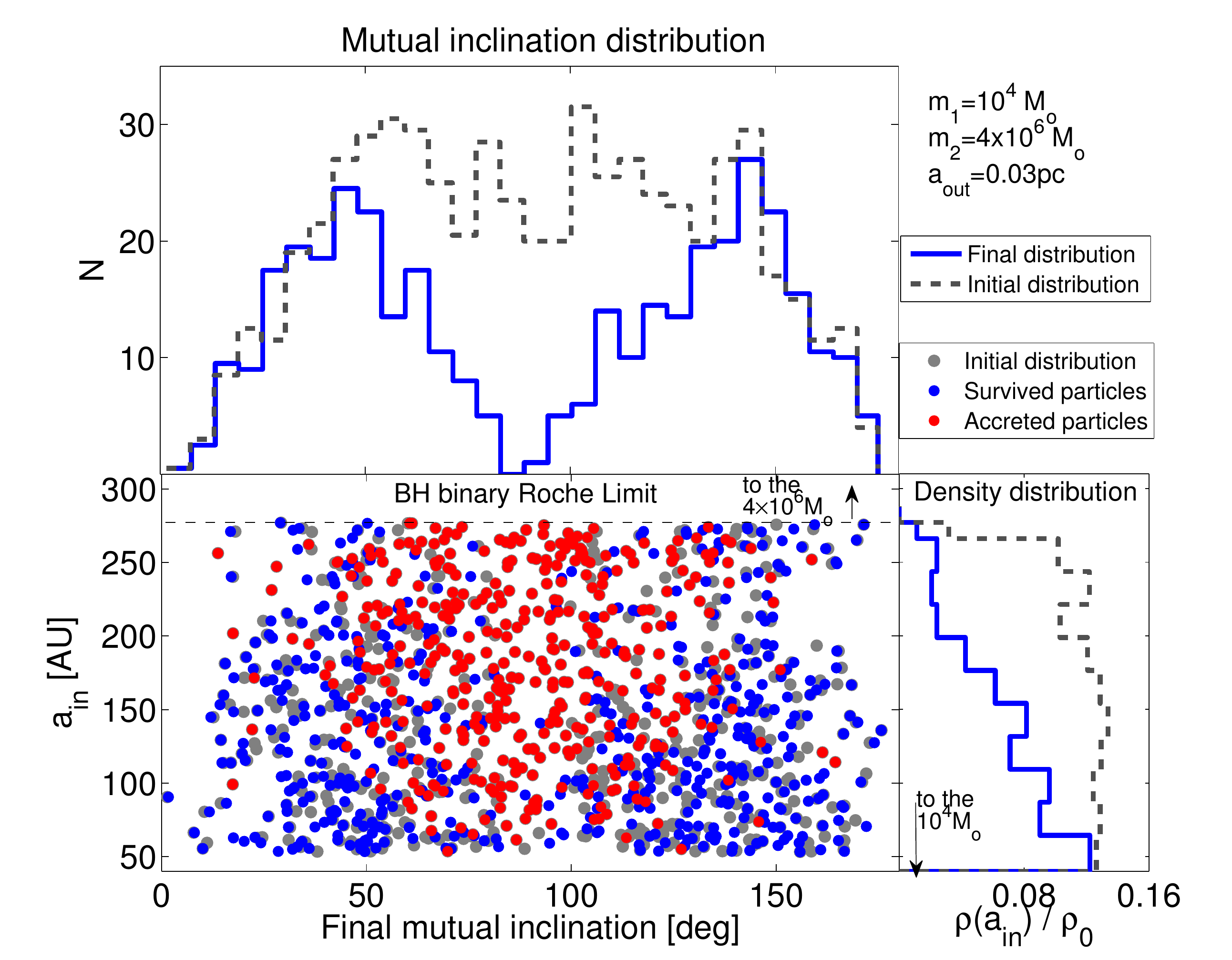}
\caption{{\bf The final DM particles distribution around IMBH with a growing MBH. } Results are shown  after the MBH grows by a  factor of $\sim2$. Same ine and color  convention as figure \ref{fig:IMBH}, $a_{out}=0.03$~pc, with MBH set initially to $10^6$~M$_\odot$. The plot shows the result of $1195$ runs.
}\vspace{0.7cm}
   \label{fig:IMBHgrow} 
\end{figure}

The MBH growth will shrink the binary BH separation and thus results in increasing $\epsilon$. The latter may produce  eccentricity excitations for larger parts of the parameter space. As shown in the  example  in Figure \ref{fig:ExMBHG},  low amplitude eccentricity and inclination oscillations (at the beginning of the evolution) can be excited to larger values as the outer orbit $a_{out}$ shrinks and $\epsilon$ rises.  As shown in this figure, the large eccentricity drives the pericenter of the  DM particle beyond the $r_c$ limit.

The growth of the perturber mass enables larger parts of the parameter space to be affected by the EKL mechanism, as depicted in Figure \ref{fig:IMBHgrow}. We show this behavior by running Monte-Carlo simulations, for an IMBH-MBH binary ($10^4$ - $10^6$~M$_\odot$, respectively)  set initially at $0.03$~pc, other initial conditions are identical to those in Figure \ref{fig:IMBH}. Figure \ref{fig:IMBHgrow} shows the result after the MBH grows by about a factor 2, where about $56\%$ of all DM particles reached the  $4 m_1 G/c^2$ limit.  The DM particles around the IMBH formed a very diluted disk. It has similar features to the torus formed around the IMBH in the static MBH case, but here the torus is confined to smaller angles, with a  significant reduction of the DM density (see right panel in Figure \ref{fig:IMBHgrow}).

% 1.000000e+04  1.000000e+06  9.792091e+01  6.187944e+03 4.921883e-03  7.000000e-01  9.013257e+01  1.550944e+01  6.680589e+01 100.

\section{DM particles around SMBH}\label{sec:SMBH}

The behavior described here is not limited to MBH-IMBH binaries. As depicted in Figure \ref{fig:timescales}, the mechanism depends on the mass ratio and the semi-major axes ratio  of the two orbits, as well as the eccentricity of the outer orbit.  We therefore expand our study to the effects of the EKL mechanism  on DM particles distribution in the case of Supermassive Black Hole (SMBH) binaries.
As mentioned above, these type of binaries are an expected result of major galaxy mergers, and their formation  has been studied extensively using hydrodynamic simulations \citep[e.g.,][]{DiMatteo+05,Hopkins+06,Robertson+06,Callegari+09}. Assuming that during a galaxy merger a SMBH binary is formed with a large mass ratio, we study the effect of the EKL mechanism on the DM particle distribution around the less massive SMBH (see Figure \ref{fig:timescales}).

\begin{figure}[!t]
\hspace{-0.7cm}
\includegraphics[width=10cm,clip=true]{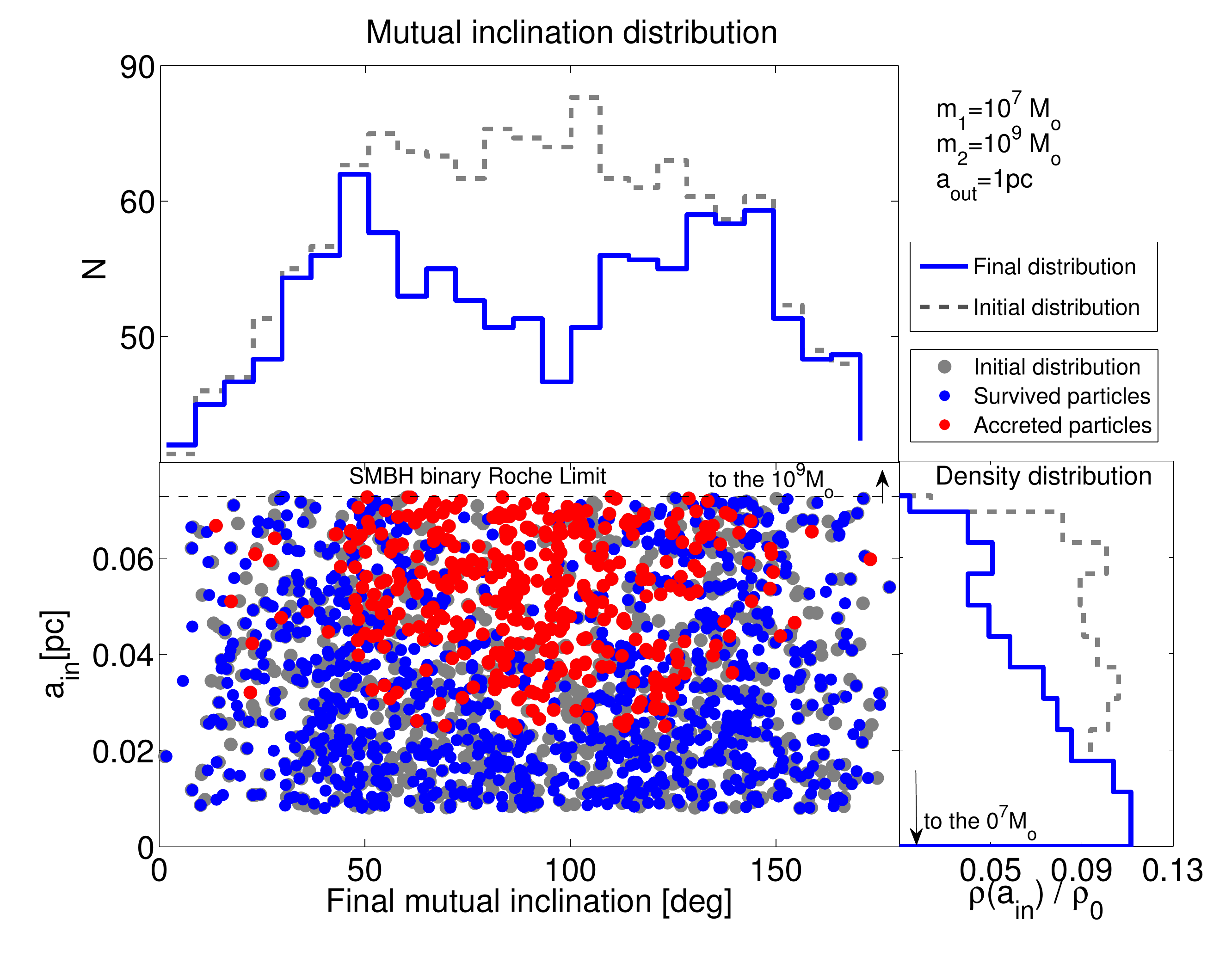}
\caption{{\bf The final DM particles distribution around SMBH after 1Gyr of evolution.} We consider $m_1=10^7$~M$_\odot$ at $1$~pc from the $m_2=10^9$~M$_\odot$. Same line and color convention as Figure \ref{fig:IMBH}. The plot shows the result of $1500$ runs. 
}\vspace{0.7cm}
   \label{fig:1e71e9} 
\end{figure}

The probability  of forming a SMBH binary with a large mass ratio is hard to estimate. It depends on the dynamical friction time-scale, and  thus the mass ratio of the host galaxy will play an important role. \citet{Khan+12} studied the formation of BH binaries through galaxy minor mergers and concluded that such systems on a tight orbit are possible.   They have looked at a mass ratio of $1:10$ and found relatively efficient SMBH inspiral. Here we have extrapolated their results to $1:100$ ratio. 
 However, the probability of forming such systems still needs to be quantified. Furthermore,  gravitational wave emission can cause black holes to recoil at escape speeds  and wander around the halo of a galaxy after the BHs merger \citep[e.g.,][]{Merritt+04,Campanelli+07,Blecha+08}. Thus they may not be present at the center of galaxies to form SMBH binaries in subsequent mergers. However this is more significant for small mass ratio SMBH binaries \citep[e.g.,][]{Lousto+12}.

In Figure \ref{fig:1e71e9} we show the result of the final distribution of DM particles around a $10^7$~M$_\odot$ SMBH in a binary configuration with $10^9$~M$_\odot$ SMB, with $a_{out}=1$~pc and $e_{out}=0.7$, after $1$~Gyr of integration time.  In this example no growth of the BH mass was assumed.  In this configuration the DM particle density was also chosen from a core distribution, i.e., uniform in separation as expected for SMBH binaries in elliptical galaxies \citep[e.g.,][]{Quinlan+97}\footnote{ We note however that a torus is formed independently of the initial DM distribution around the BH. Thus, even if these galaxies had a cusp-like DM distribution initially \citep[as suggested by, e.g.,][]{Lauer+95,Lauer+07}, the  result of a final torus-like configuration is preserved.}.  The arguments of the pericenters and longitudes of ascending nodes were chosen  from a uniform distribution. We choose an isotropic distribution for the DM particles inclinations and a uniform distribution for their eccentricity.  Similarly to the examples showed in \S \ref{sec:IMBH}, the upper limit inner orbit separation was chosen to be inside the Roche limit, which implies $\epsilon<0.09$.
We note that the Newtonian precession is comparable to $t_{oct}$ for densities of about $10^7$~M~pc$^{-3}$ in the inner parts ($\lsim 0.08$~pc) of the $10^7$~M$_\odot$ BH.

About {\bf $29\%$} of all DM particles have accreted onto the SMBH due to the EKL mechanism. DM particles with $\epsilon < 0.034$ (corresponding to $a_{in}<0.025$~pc) did not accrete onto the SMBH (see figure  \ref{fig:1e71e9} bottom panels). Therefore an incomplete torus was formed.  However, particles with extremely small pericenters could mostly survive in low inclination orbit configurations. Of course similarly to the example shown in section  \ref{sec:growMBH}, growth of the BHs as a function of time can allow larger parts of the parameter space to be affected by the EKL mechanism. 

Note that stars in a bulge mass of $\sim 10^7$~M$_\odot$ can scatter the  DM particles \citep{Vasiliev+08}. We assume a uniform eccentricity distribution, which may account for this effect. However if the scattering takes place on much shorter time-scales then the secular timescale, it can change the DM particle separations and thus our secular approximation may not be valid.

 \section{Discussion}\label{sec:Dis}
 
 We have studied the  dynamical evolution of DM particles around BH binaries. In this hierarchical configuration, a DM particle and the BH are considered as an inner binary. The companion BH, on a much wider orbit, causes gravitational perturbations on the inner binary. We showed that this secular evolution (and specifically the  {\it eccentric Kozai--Lidov} (EKL) mechanism) plays an important role in redistributing the DM particles around the {\it less} massive member of the BH binary.    The EKL mechanism can drive the DM particles that are  initially in a near-polar configuration  to very large eccentricities that may result in their accretion onto the BH. Depletion near the polar axis yields a torus-like configuration.

 We showed that due to stability and time-scale requirements, DM particles around the less massive BH binary member will be affected more from the EKL mechanism (see Figure   \ref{fig:timescales}). We considered two main  representative  examples; an IMBH-MBH binary and a SMBH binary. 
 For the MBH-IMBH binary we assume an IMBH in the vicinity of the galactic center and we explore two scenarios:  $a_{out}=0.03$~pc  and $a_{out}=1$~pc depicted in Figures \ref{fig:IMBH} and \ref{fig:IMBH1pc}.

DM particles will accrete onto the IMBH  if the pericenter distance can reach extremely large values, which we chose to be $r_c=4m_1 G/c^2$, following  \citet{GS99}.  For an IMBH with $10^4$~M$_\odot$, this limit is $r_c=3.95\times 10^{-4}$~AU ($=1.92\times 10^{-9}$~pc) which corresponds to extremely high eccentricities.  In the $a_{out}=0.03$~pc example this eccentricity ranges from  $1-e_{in}\sim 2\times 10^{-6}$ to $10^{-5}$. This range corresponds to the DM separation distribution around the IMBH, with an upper limit satisfied by the IMBH Roche limit [Eq.~\ref{eq:Roche}], and a lower limit for which $t_{quad}\sim t_{GR}$, respectively. For the $a_{out}=1$~pc example, the eccentricity that DM particles need to reach to cross the $r_c$ limit ranges between $1-e_{in}\sim6.8\times 10^{-8}$ to $7.6\times 10^{-7}$, respectively.  The EKL mechanism is very efficient in producing large eccentricities associated with the flip of the orbit \cite[e.g.][]{Naoz11}, but the eccentricity does not necessarily reach unity  \citep[e.g.,][]{Li+13}. Furthermore, configurations that reach eccentricities with $1-e_{in}<10^{-7}$ are harder to achieve (see for detailed analysis in the test particle approximation \citet{Li+14}). These differences are depicted in Figure \ref{fig:IMBHmaxe}, which explains why the $a_{out}=0.03$~pc  example  shows a larger  effect due to the EKL mechanism. Most of the near-polar configurations reached large eccentricities (which represent  about $21\%$ of all the runs) on a typical time-scale of $0.3$~Myr. The final DM inclination distribution deviates from spherical  symmetry and produces a  double peak distribution  with preference at $40^\circ$ and $140^\circ$ (Figures \ref{fig:IMBH}, top panel). This yields a torus-like configuration with a diluted DM density further away from the IMBH (as depicted in the cartoon in Figure \ref{fig:config}).

We also considered the effect of an adiabatic growth of the MBH (the perturber in our case)  on the DM distribution around the IMBH located at $a_{out}=0.03$~pc from the MBH (Figures \ref{fig:ExMBHG} and \ref{fig:IMBHgrow}). 
A growing MBH shrinks the separation between the IMBH and MBH, and thus increases the efficiency of the EKL mechanism, as it can tap into larger parts of the parameter space. This has resulted in $\sim 56\%$ accreted particles, and a much thiner torus (see Figure \ref{fig:IMBHgrow}).

\begin{figure}[!t]
\hspace{-0.7cm}
\includegraphics[width=8.5cm,clip=true]{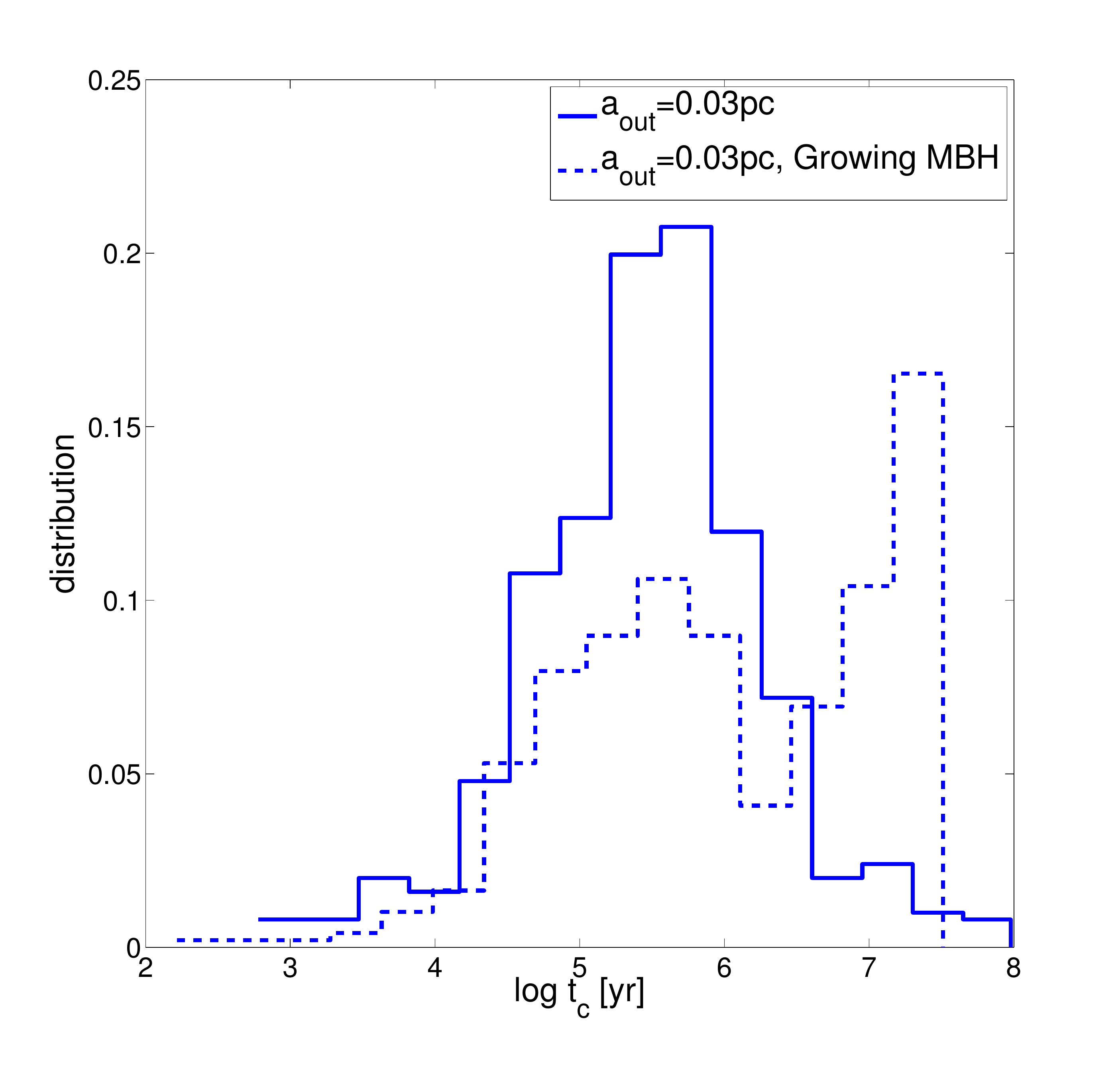}
\caption{{\bf The time distribution of accreted DM particles}.  We consider the IMBH-MBH example setting $a_{out}=0.03$, without (solid line) and with (dashed line) growing MBH. The distribution is normalized  such that the integral of the  distribution is unity.
}\vspace{0.7cm}
   \label{fig:tDis} 
\end{figure}

We suggest that these results may be relevant to indirect DM signatures. If the IMBH is spinning, than these particles arriving into the ergosphere will  have an enhanced chance to interact and 
self-annihilate, with debris emission possibly boosted by the Penrose effect \citep{Banados+11,Zaslavskii12}.% \cite{}.

Note that the initial DM distribution around the IMBH does not affect the formation of a torus. It may be significant for studies of the   distribution of the DM particles inside the torus or the flux at which DM particles arrive at the IMBH. Therefore, although  we have assumed a core distribution around the IMBHs,  the result can be easily generalized to other DM distributions. 

 In Figure \ref{fig:tDis} we show the time distribution of crossing $r_c$ for the $a_{out}=0.03$~pc case, with and without an adiabatically growing MBH.  When the MBH does not grow in mass, the EKL mechanism works efficiently in the first few to ten $10^5$~yrs (as apparent by the same shape distribution of  the two cases). The MBH grows on a Salpeter growth time scale, and already when the MBH grows to about $10\%$ from its original value, $\epsilon$ increases as well in about the same amount (and continued to increase). This is enough to re-trigger the EKL mechanism for larger parts of the parameter space (as depicted in Figure \ref{fig:tDis}). 
Interestingly, the recent interpretations of \citep{Daylan+14} of the gamma ray signal from the galactic center as an evidence for annihilating DM, located at about $1^\circ$ on the sky, may also arise due to the presence of dark matter annihilations enhanced around IMBHs close to the central SMBH \citep[see also][]{Finkbeiner+14}. 

We  have also expanded our study and  applied this mechanism to SMBH binaries, again focusing on the DM distribution around the less massive member (see Figure \ref{fig:1e71e9}). We considered a  $10^7$~M$_\odot$-$10^9$~M$_\odot$ binary separated at $1$~pc and showed that the near-polar outer parts of the DM sphere around the $10^7$~M$_\odot$ SMBH will be depleted, yielding a torus-like configuration which is diluted further away from the SMBH. Similarly to the MBH-IMBH DM annihilations arguments, if the SMBH spins, we would expect the DM particles to linger on the ergosphere allowing for the possible occurrence of enhanced and energetic (in centre-of-mass)  self annihilation processes. 

Finally we note that torus-like configurations of dark matter around a SMBH or an IMBH may be a optional source for  gravitational wave signals  \citep{Eda+13}\footnote{Note that \citet{Eda+13} proposed the existence of a spike around the IMBH, while we have assumed a core. However, as mentioned above, the formation of a DM torus does not depend on the initial DM distribution.  } and  gravitational lensing enhancements of the IMBH shadow \citep{Inoue+13}. Furthermore, we suggest that  the DM torus in combination with the EKL mechanism may have interesting implications for stellar capture as well as  dynamical friction of stars, which is highly depended  on the density of the DM torus.  More detailed calculations are needed to assess these issues.

\section*{Acknowledgments}
We thank  Avi Loeb, Caroline Terquem, Laura Blecha, Doug Finkbeiner and Enrico Barausse  for  useful discussions. We also thank the anonymous referee for carefully reading the manuscript and providing valuable suggestions. SN was partly   supported by NASA through a Einstein Post--doctoral Fellowship awarded by the Chandra X-ray Center, which is operated by the Smithsonian Astrophysical Observatory for NASA under contract PF2-130096. The research of JS has been supported at IAP by  the ERC project  267117 (DARK) hosted by Universit\'e Pierre et Marie Curie - Paris 6   and at JHU by NSF grant
OIA-1124403.

\bibliographystyle{hapj}

\bibliography{Kozai}

\end{document}